# Super strong paramagnetism induced by polar functional groups and water

Jihong Wang,[1†] Yizhou Yang,[1†] Jie Jiang,[2†] Liuhua Mu,[2] Guosheng Shi,[3] Yongshun Song,[1] Haijun Yang,[4] Peng Xiu,[5*] Haiping Fang[1*]

1. School of Science, East China University of Science and Technology, Shanghai 200237, China
2. CAS Key Laboratory of Interfacial Physics and Technology, Shanghai Institute of Applied Physics, Chinese Academy of Sciences, Shanghai 201800, China
3. Shanghai Applied Radiation Institute, Shanghai University, Shanghai 200444, China
4. Interdisciplinary Research Center, Shanghai Synchrotron Radiation Facility, Zhangjiang Laboratory (SSRF, ZJLab), Shanghai Advanced Research Institute, Chinese Academy of Sciences, Shanghai 201204, China
5. Department of Engineering Mechanics, Zhejiang University, Hangzhou 310027, China

[†]These authors contributed equally to this work.
[*]Corresponding authors. Email: fanghaiping@sinap.ac.cn; xiupeng2011@zju.edu.cn

**Abstract:** We experimentally demonstrate that some commonly used materials such as cellulose acetate and chitin − which are traditionally considered to be non-magnetic − show super strong paramagnetism in aqueous solutions under ambient conditions when they are agglomerated by nanoparticles. Theoretical computations show that strongly polar functional groups can reduce the potential barrier for a singlet-triplet interconversion with the help of surrounding water, inducing the magnetic moments. These magnetic moments distributed on the surfaces of the nanoparticles, which make a large number of magnetic moments gather in a very small space, greatly enhance the alignment of the moments along and amplify the effect of the external magnetic field, resulting in the super strong paramagnetism. Our findings suggest that the polar functional group may always induce paramagnetism and the magnetic effect may be universal as the electric effects since the polarization is very common in materials.

Conventionally, magnetic materials comprise some metals (e.g., Fe, Co, and Ni), metal oxides, intermetallic compounds, and organic/molecule-based materials with radicals (*1-4*). Recently, some nanostructured carbon materials such as graphene have recently been reported to exhibit room-temperature ferromagnetism due to the edge/defect effects (*5*). Very recently, we have reported the ferromagnetic properties of two-dimensional CaCl crystals in graphene membranes (*6*), and super strong paramagnetism on aromatic peptides induced by cations (*7, 8*), which are attributable to the valent modification of the cations, due to the cation-π interaction between cations and aromatic rings.

In this paper, we show that, without all the traditional materials and metal cations, super strong paramagnetism can be observed. Explicitly, we experimentally demonstrate that some commonly used materials such as cellulose acetate and chitin show strong paramagnetism in aqueous solutions under ambient conditions, when they are agglomerated by nanoparticles. Theoretical computations show that both the strongly polar functional groups and water can substantially reduce the energy difference between the singlet and triplet states of the cellulose acetate monomer which contains one glucose monomer with two ethyl acetate groups, inducing the magnetic moment due to the existence of the triplet states. The magnetic moments on the surfaces of nanoparticles gather in a very small space greatly enhance the alignment of the moments and amplify the effect of the external magnetic field, thus resulting in the super strong paramagnetism.

In the experiment, the powders of cellulose acetate (Acetyl 39.8 wt%, hydroxyl 3.5 wt%) were added into a plastic tube containing pure water (Milli-Q, 18.2 MΩ), shaken well to prepare cellulose acetate aqueous dispersion, and stored still at 25 °C. After settling for 30 minutes, the bottom solution was taken out into a cuvette and six times more pure water was added. Then the cuvette was gently shaken to uniformly disperse the samples, and settled for 1 minutes.

We placed a magnet to the side of a cuvette. The magnet is a commercial Nd-Fe-B magnet with a surface magnetic field intensity of ~0.5 T. Fig. 1 shows some typical snapshots at different time points. Initially, the insoluble cellulose acetate particles dispersed in water when a magnet was just placed close to the cuvette. At the time $t = $ 10 minutes, many particles are attracted towards the magnet and accumulated near the magnet. We then moved the magnet to the right, it can be seen clearly that the adsorbed particles followed the motion of the magnet accordingly (see the last three snapshots in Fig. 1). These results indicate that cellulose acetate nanoparticles are attracted and dragged by a magnet nearby. We note that the magnetic field intensity of the magnet is only ~0.5 T thus showing that the cellulose acetate particles have super strong paramagnetism in aqueous solutions.

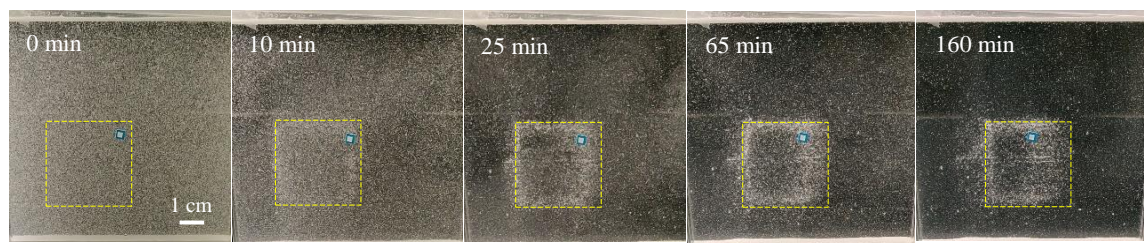

**Figure 1.** Snapshots at different time points to show the attraction and dragging of cellulose acetate particles in aqueous solutions by a regular magnet. The yellow dashed square indicates the position of the magnet, and the tiny white dots represents

the cellulose acetate particles. The small hollow blue square indicates a fixed position on the cuvette for guiding the eye.

Next, we performed density functional theory (DFT) calculations to explore the underlying physics of the observed super strong paramagnetism. Cellulose acetate is a derivative of cellulose. One cellulose acetate molecule is usually composed of at least thousands of monomers; each typical cellulose acetate monomer contains one glucose monomer with two hydroxyl groups substituted by the ethyl acetate groups. A cellulose acetate monomer usually has a cyclic structure of the glucose which would occasionally transform into the open-ring form in aqueous solutions. We focus on two states of the monomer of cellulose acetate – i) the singlet state with a cyclic form, without unpaired electron spins and the magnetic moment; and ii) the triplet state with an open-ring form, with unpaired electron spins, carrying the magnetic moment. The DFT calculations shows the singlet state is much more stable than the triplet one, with the energy difference between two states ~56 kcal/mol. We note that for the cellulose monomer, without the ethyl acetate group on the side chain, the energy difference between singlet and triplet states is much larger, reaching ~87 kcal/mol, suggesting that the electric polarization of the functional group can greatly reduce the potential barrier in the singlet-triplet interconversion.

Water also plays a key role in stabilizing the triplet state. We introduce three surrounding water molecules into the system. DFT calculations show that the energy difference between singlet (Fig. 2a) and triple states (Fig. 2b) dramatically decreases to ~18 kcal/mol which is ~3.3 times of the hydrogen-bond energy of liquid water. We expect such value can be reduced by further optimizing the structure so that the energy barrier for the singlet-triplet interconversion can be occasionally overcome by thermal fluctuations in aqueous solutions. Therefore, the strongly polar carboxyl group together with water can greatly facilitate the excitation of some cellulose acetate monomers from the singlet state with a cyclic form into the triplet state with an open-ring form, thus inducing the magnetic moments.

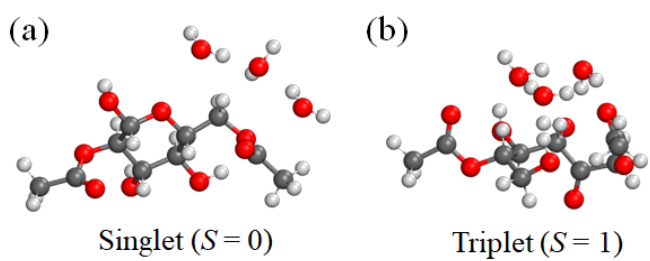

**Figure 2.** Optimized structures of the cellulose acetate monomer together with three water molecules obtained from density functional theory (DFT) calculations. (a) the singlet state ($S = 0$) with a cyclic form; (b) the triplet state ($S = 1$) with an open-ring form. Spheres in grey, red, and white represent carbon, oxygen, and hydrogen, respectively.

We all know that the interaction is extremely weak between a single magnetic moment and an external magnetic field (*9*). Thus, even some monomers of the cellulose acetate nanoparticles with triple states possess net magnetic moments, how can they show an observably strong paramagnetism? The key is the size of the nanoparticles. When a huge number of monomers with triple states distributed on the surfaces of the nanoparticles, they gather in a very small space. These monomers are close enough so that an additional magnetic field produced by any monomer will acts on all the other monomers, and the magnetic fields of other monomers would in turn act on this monomer itself. This positive-feedback greatly enhances the alignment of the magnetic moments of those monomers, thus amplifying the effect of the external magnetic field and resulting in the strong paramagnetism of the cellulose acetate nanoparticles.

Now we understand that the cooperative action of the strongly polar functional groups – carboxyl, and water, greatly reduce the energy barrier of the singlet-triplet interconversion of the cellulose acetate monomer, and the magnetic moments due to those triplet states on the surfaces of nanoparticles gathering in a very small space, greatly amplify the effect of the external magnetic field, resulting in the super strong paramagnetism of those nanoparticles. We note that neither the dry cellulose acetate powders nor the precipitations (nanoparticles assembling into particles of larger sizes) on the bottom of the cuvette can be attracted by the magnet, showing that they do not have strong paramagnetism. Clearly, both of those states do not have enough water molecules involved.

We perform additional experiments in aqueous solutions using nanoparticles of various cellulose-based materials, including chitin, chitosan, and cellulose acetate butyrate. Chitin particles can be attracted and dragged by the magnet, exhibiting super strong paramagnetism in aqueous solutions. Other materials can also be clearly attracted by the magnet. All of these materials have polar functional groups on the side chain, further support the mechanism presented here.

The strong paramagnetism is not limited to the cellulose-based molecule. Experimentally, we found that the 2-Naphththyl acetate consisting of an ethyl acetate group and an aromatic naphthalene ring, can be attracted by the magnet as well in aqueous solutions. Similar to cellulose acetate, 2-Naphththyl acetate also has a carboxyl group. We thus expect that magnetic moments can be always induced by polar functional groups. Generally, super strong paramagnetism can only be induced by the strongly polar functional group.

To summarize, we have experimentally demonstrated that some commonly used materials such as cellulose acetate, chitin, chitosan, cellulose acetate butyrate and 2-Naphththyl acetate in aqueous solutions under ambient conditions, when they are agglomerated by nanoparticles, display strong paramagnetism and some even display super strong paramagnetism. DFT calculations show that the cooperative interaction of the strongly polar functional group and water greatly reduces the energy barrier of the singlet-triplet interconversion on the monomer of cellulose acetate, thus inducing the magnetic moments on the surface of nanoparticles.

It seems that the existence of the magnetic moments is quite universal, no matter it is weak or strong, at the interface between water and the molecules with polar functional groups. One may ask a question: why the magnetic effects have not been reported before although those interfaces are very common? They key is the nanoparticles. Since the magnetic moments only exist at the interfaces, for particles with enough large sizes, magnetic moments are only distributed on the two-dimensional surfaces of the particles. There are not enough magnetic moments gathering in a very small space that the additional magnetic fields can interacts between them. On the other hands, when the particles are small enough, the magnetic moments on all or most of the surfaces of a nanoparticle is very close so that an additional magnetic field produced by any of the magnetic moments can act on all the magnetic moments on the surface of the nanoparticle, which greatly enhance the alignment of the moments and amplify the effect of the external magnetic field, resulting in the observably strong paramagnetism. This is correlated with our previous observation on the ferromagnetic properties of two-dimensional CaCl crystals in the graphene membrane (*6*), where the membrane provides enough surfaces within a very small space.

Finally, we would like to see that water may be not the only material to help the polar functional groups induce the magnetic moments. The strong paramagnetism maybe universal for materials including functional groups with sufficiently strong polarity. Our findings open new perspective on the field of magnetism and are expected to have wide technological applications.

**References:**
1. P. Perlepe *et al.*, Metal-organic magnets with large coercivity and ordering temperatures up to 242°C. *Science* **370**, 587-591 (2020).
2. J. S. Miller, Magnetically ordered molecule-based materials. *Chem. Soc. Rev.* **40**, 3266-3296 (2011).
3. J. S. Miller, Organic- and molecule-based magnets. *Mater. Today* **17**, 224-235 (2014).
4. S. Venkateswarlu, M. Yoon, New Room-Temperature Organic Molecule-Based Magnets. *Trends in Chemistry* **1**, 363-364 (2019).
5. G. Z. Magda *et al.*, Room-temperature magnetic order on zigzag edges of narrow graphene nanoribbons. *Nature* **514**, 608-611 (2014).
6. L. Zhang *et al.*, Novel 2D CaCl crystals with metallicity, room-temperature ferromagnetism, heterojunction, piezoelectricity-like property, and monovalent calcium ions. *National Science Review*, nwaa274, doi:210.1093/nsr/nwaa1274 (2020).
7. H. Yang *et al.*, Unexpectedly super strong paramagnetism of aromatic peptides due to cations of divalent metals. *arXiv:2011.12797*, cond-mat.soft (2020).
8. S. Sheng *et al.*, Super strong paramagnetism of aromatic peptides adsorbed with

monovalent cations. *arXiv:2012.11850*, Physics.bio-ph (2020).
9. S. Blundell, *Magnetism in Condensed Matter*. Oxford University Press (2001).